\documentclass[preprint,superbib,byrevtex,showpacs,prb]{revtex4}%
\usepackage{amsfonts}
\usepackage{amsmath}
\usepackage{amssymb}%
\usepackage{graphicx}
\newtheorem{theorem}{Theorem}
\newtheorem{acknowledgement}[theorem]{Acknowledgement}

\begin{document}
\title{A MODEL FOR NONEXPONENTIAL RELAXATION AND AGING IN DISSIPATIVE SYSTEMS}
\author{A. P\'{e}rez-Madrid}
\affiliation{Departament de F\'{\i}sica Fonamental.
Facultat de F\'{\i}sica. Universitat de Barcelona.
Diagonal 647, 08028 Barcelona. Spain}

\keywords{one two three}
\pacs{64.70.Pf, 05.10.Gg, 05.40.-a}

\begin{abstract}
The nonexponential relaxation and aging inherent to complex dynamics
manifested in a wide variety of dissipative systems is analyzed through a
model of diffusion in phase space in the presence of a nonconservative force.
The action of this force establishes a heat flow which maintains the system
away from equilibrium. The inability of the system to find its equilibrium
state becomes apparent through the presence of an effective temperature field.
This is the temperature of the stationary nonequilibrium state reached by the
system satisfying a generalyzed version of the fluctuation-dissipation
theorem. \ The presence of a nonequilibrium temperature leads to a hierarchy
of relaxation times responsible for the aging phenomena and to a relation
similar to the Vogel-Fulcher-Tammann law.

\end{abstract}
\date[10/03/04]{date}
\maketitle

\section{Introduction}

Under certain conditions dissipative systems give rise to an often complex
dynamics. This is the case of glassy systems which constitute an ubiquitous
example of dissipative systems following a complex dynamics \cite{ngai}. Glass
is a nonequilibrium system whose dynamics is nonlinear and has nonexponential
relaxation functions which leads to the aging phenomena. Similar to glass is
the process of protein folding which involves crossing a large number of
energy barriers \cite{wolynes1}, \cite{sabelko}. \ Anomalous transport
processes and strange kinetics based on fractional generalizations of the
diffusion and Fokker-Planck equations \cite{shlesinger}, \cite{balescu}
constitute other typical examples. Hence, complex dynamics is the object of
active research due to its multiplicity of manifestations. Here, we will focus
on the nonexponential character of the relaxation functions and on aging in
complex dynamics. The mechanisms underlying slow relaxation and aging still
lack a clear and definitive elucidation. In glasses and proteins for example,
several experiments and computer simulations have been done which support the
explanation of these relaxation phenomena in the framework of the energy
landscape paradigm as the result of activated diffusion through a rough energy
landscape of valleys and peaks \cite{stillinger}-\cite{ediger}.

To contribute to the understanding of what these mechanisms are, we propose
here a simple model to show a possible origin of nonexponential relaxation and
aging based on the idea of the energy landscape and nontrivial energy
barriers. This model which consists of the diffusion in phase space, provides
a direct link between the phase space dynamics and the slow relaxation of the
functions of the configuration of the system in the corresponding energy
landscape. The slowing down of the dynamics clearly appears as a consequence
of the freezing of some degrees of freedom which takes the system out of
equilibrium. This fact is indicated by the presence of an effective
temperature field incorporating the information of the suppressed degrees of
freedom and depending on the equilibrium temperature at the moment the quench
was applied.

The paper is distributed as follows. In the second section we develop the
thermodynamic framework in which our model is embedded. In section three, we
write the Fokker-Planck equation governing the dynamics and study the
relaxation. Section four is devoted to deriving the generalized diffusion
equation and obtaining the relaxation function and the hierarchy of relaxation
times in some particular cases. Section five deals with the computation of the
generalized fluctuation-dissipation theorem GFDT. Finally in section seven, we
discuss our main conclusions.

\section{Thermodynamic framework}

We model the nonexponential relaxation phenomenon in a dissipative system as
the Brownian motion of a particle of unit mass in a nonperiodic potential
$V(x)$, able to constitute a schematic representation of a rough energy
landscape. Moreover, we will apply the methods of nonequilibrium
thermodynamics at a mesoscopic level which consists of computing the entropy
production from the Gibbs entropy postulate and postulating the linear
phenomenological equations relating the fluxes and the thermodynamic forces
occurring therein. This constitutes a mesoscopic generalization of the
macroscopic formalism of the nonequilibrium thermodynamics, the mesoscopic
nonequilibrium thermodynamics (MNET)\cite{inertial}, \cite{mesoscopic}.

Thus, our thermodynamic study arises from the assumption that we are dealing
with a physical ensemble of Brownian particles, a thermodynamic system to
which thermodynamic relations can be applied \cite{hill}. We consider that the
ensemble is initially at equilibrium in a heat bath at temperature $T_{o}$,
and thus, is distributed according to the canonical distribution
\begin{equation}
\psi_{eq}\sim\exp\left[  -\frac{H(\Gamma)}{k_{B}T_{o}}\right]  \text{ , }
\label{canonical}%
\end{equation}
where $H(\Gamma)=$ $1/2u^{2}+V(x)$ is the Hamiltonian, and $\Gamma=(x,u)$
represents a point of the one-particle phase space, with $x$ being the
position and $u$ the velocity. At a certain moment ($t=0$), we apply a
nonconservative force $f(t)$ quenching the system. It is assumed that the
nonequilibrium entropy of the system is given through the Gibbs entropy
postulate \cite{vankampen}, \cite{degroot}
\begin{equation}
S(t)=-k\int\psi\ln\frac{\psi}{\psi_{eq}}d\Gamma+S_{eq}\text{ ,}
\label{gibbsentropy}%
\end{equation}
where $S_{eq}$ is the equilibrium entropy. Variations in the probability
density $\psi(\Gamma,t)$ imply changes in the entropy which can be obtained
from Eq. ( \ref{gibbsentropy}). These are given by%
\begin{equation}
\delta S=-\int\left(  k\ln\frac{\psi}{\psi_{eq}}+\frac{\mu_{eq}}{T_{o}%
}\right)  \delta\psi d\Gamma\text{ ,} \label{entropychange}%
\end{equation}
where we have taken into account that $\delta S_{eq}=-\int\mu_{eq}/T_{o}%
\delta\psi d\Gamma$, with $\mu_{eq}=-V(x)+\mu_{o}$ being the equilibrium
chemical potential per particle of the Brownian gas and $\mu_{o}$ a constant .
On the other hand, we assume local equilibrium in the phase space $\Gamma$
consisting in postulating the Gibbs equation at a local level in the phase
space \cite{inertial}, \cite{mesoscopic}, which in the case of a homogeneous
bath, reduces to%
\begin{equation}
T_{o}\delta S=-%
{\displaystyle\int}
\mu(\Gamma,t)\delta\psi(\Gamma,t)d\Gamma\text{ ,} \label{gibbs}%
\end{equation}
where the thermodynamically conjugated of the density $\psi(\Gamma,t)$
nonequilibrium chemical potential $\mu(\Gamma,t)$ appears. The expression of
the nonequilibrium chemical potential is obtained by comparing Eqs.
(\ref{entropychange}) and (\ref{gibbs}), thus
\begin{equation}
\mu(\Gamma,t)=kT_{o}\ln\frac{\psi}{\psi_{eq}}+\mu_{eq}\text{ .}
\label{chemical potential}%
\end{equation}

A gradient of the chemical potential induces a diffusion process in phase
space in order to restore the equilibrium state. Throughout this process, the
distribution of probability changes according to the following Generalized
Liouville equation%
\begin{equation}
\frac{\partial}{\partial t}\psi(\Gamma,t)+V_{\Gamma}(\Gamma,t)\cdot
\nabla_{\Gamma}\psi(\Gamma,t)=-\frac{\partial}{\partial u}\psi(\Gamma
,t)f(t)-\frac{\partial}{\partial u}J(\Gamma,t)\text{ ,} \label{Liouville}%
\end{equation}
where $V_{\Gamma}(\Gamma,t)=\left(  u,-\nabla V(x)\right)  $ is the velocity
corresponding to the Hamiltonian flow, $J(\Gamma,t)$ constitutes a diffusion
current and $\nabla_{\Gamma}=\left(  \frac{\partial}{\partial u}%
,\nabla\right)  $, where $\nabla$ stands for the spatial derivative. In order
to completely characterize the diffusion process, one has to find the
expression of the diffusion current by following the rules of nonequilibrium
thermodynamics \cite{degroot}. In this framework, through the computation of
the entropy production one can establish phenomenological relations between
currents and conjugated thermodynamic forces. Thus, from Eq. (\ref{gibbs}) we
can determine the rate of change of the entropy%
\begin{equation}
T_{o}\frac{dS}{dt}=-%
{\displaystyle\int}
\mu(\Gamma,t)\frac{\partial}{\partial t}\psi(\Gamma,t)d\Gamma\text{ ,}
\label{entropy rate}%
\end{equation}
which after substituting Eq. (\ref{Liouville}) and by using Eq.
(\ref{chemical potential}) becomes
\begin{equation}
T_{o}\frac{dS}{dt}=\left\langle \left(  f(t)-\nabla V(x)\right)
u\right\rangle -%
{\displaystyle\int}
J(\Gamma,t)\frac{\partial}{\partial u}\mu(\Gamma,t)d\Gamma\text{ ,}
\label{entropyvariation}%
\end{equation}
obtained after partial integration. Here,
\begin{equation}
\left\langle \left(  f(t)-\nabla V(x)\right)  u\right\rangle =%
{\displaystyle\int}
\left(  f(t)-\nabla V(x)\right)  u\psi(\Gamma,t)d\Gamma\text{ } \label{power}%
\end{equation}
is the power supplied by the net force which is dissipated in the system as
heat. Thus, the first term on the right hand side of Eq.
(\ref{entropyvariation}) can be interpreted as the rate of heat exchange with
the surroundings
\begin{equation}
\frac{dq}{dt}=-\left\langle \left(  f(t)-\nabla V(x)\right)  u\right\rangle
\text{ , } \label{heatrate}%
\end{equation}
with $dq$ being the amount of heat released in a time $dt$. However, the
second term on the right hand side of Eq. (\ref{entropyvariation}) constitutes
the entropy production $\sigma\geq0$, according to the second law, containing
the current and its conjugated thermodynamic force $(\partial/\partial
u)\mu(\Gamma,t)$, which by postulate of the nonequilibrium thermodynamics are
related through the phenomenological law%
\begin{equation}
J(\Gamma,t)=-\frac{L}{T_{o}}\frac{\partial}{\partial u}\mu(\Gamma,t)\text{ ,}
\label{phenomenologicallaw}%
\end{equation}
where $L$ is a phenomenological coefficient. Therefore, Eq.
(\ref{entropyvariation}) may be rewritten in a more significant form
\begin{equation}
\frac{dS}{dt}=\frac{1}{T_{o}}\frac{dq}{dt}+\sigma\text{ ,}
\label{entropybalance}%
\end{equation}
expressing the entropy balance between the exchange of heat with the
surroundings and the entropy generated in the irreversible processes
established in the system when it is removed from its equilibrium state. By
using the expression of the nonequilibrium chemical potential
(\ref{chemical potential}) in (\ref{phenomenologicallaw}), one obtains%
\begin{equation}
J(\Gamma,t)=-\gamma\left(  k_{B}T_{o}\frac{\partial}{\partial u}+u\right)
\psi\text{ ,} \label{explicitcurrent}%
\end{equation}
where we have performed the identification $L/\psi T_{o}=\gamma,$ with
$\gamma$ being the friction coefficient of the Brownian particle so that%
\begin{equation}
\sigma=\frac{\gamma}{T_{o}}%
{\displaystyle\int}
\frac{\left(  k_{B}T_{o}(\partial/\partial u)\psi(\Gamma,t)+\psi
(\Gamma,t)u\right)  ^{2}}{\psi(\Gamma,t)}d\Gamma\text{ .}
\label{entropyproduction}%
\end{equation}
One can conclude that is possible the existence of a stationary nonequilibrium
state ($dS/dt=0$) in the system, \textit{i.e. }not given by an equilibrium
maxwellian, for which%
\begin{equation}
\frac{1}{T_{o}}\frac{d_{st}q}{dt}\!\,{}=-\sigma_{st}\text{ .}
\label{stationarity}%
\end{equation}

\section{Fokker-Planck equation and relaxation}

The Fokker-Planck equation governs the dissipative dynamics when the
description of the state of the system is given in terms of the probability
density $\psi(\Gamma,t)$. This equation is obtained after substituting the
expression of the current (\ref{explicitcurrent}) into Eq. (\ref{Liouville})
leading to%
\begin{gather}
\frac{\partial}{\partial t}\psi(\Gamma,t)=-\frac{\partial}{\partial x}%
u\psi(\Gamma,t)+\frac{\partial}{\partial u}\psi(\Gamma,t)\left(  \nabla
V(x)-f(t)\right) \label{F-P}\\
+\gamma\frac{\partial}{\partial u}\left(  \beta^{-1}\frac{\partial}{\partial
u}+u\right)  \psi(\Gamma,t)\text{ .}\nonumber
\end{gather}
We will assume that $u$ is the fast variable, thus the dynamical processes in
the system are associated to configurational changes related to $x$, the slow
variable. Thus, we write%

\begin{equation}
\psi(\Gamma,t)=\phi_{x}(u,t)\rho(x,t)\text{ ,} \label{factorization}%
\end{equation}
where $\phi_{x}(u,t)$ is the conditional probability . \ The probability
density is $\rho(x,t)=\int\psi(\Gamma,t)du$ evolves according to
\begin{equation}
\frac{\partial}{\partial t}\rho(x,t)=-\nabla\int u\psi(\Gamma,t)du\text{ ,}
\label{continuity}%
\end{equation}
obtained by partial integration of Eq. (\ref{F-P}), which defines the current
$J(x,t)=\int u\psi(\Gamma,t)du.$ This current satisfies the equation
\begin{gather}
\frac{\partial}{\partial t}J(x,t)+\gamma J(x,t)=\label{precurrent}\\
-\left\{  \rho(x,t)\left(  \nabla V(x)-f(t)\right)  +k_{B}\nabla
\rho(x,t)T(x,t)\right\}  \text{ .}\nonumber
\end{gather}
which can be obtained after multiplying Eq. (\ref{F-P}) by the velocity $u$
and performing the partial integration, by using the decoupling
(\ref{factorization}). Here, $k_{B}T(x,t)=\int u^{2}\phi_{x}(u,t)du$ is the
second moment of the conditional distribution $\phi_{x}(u,t)$, playing the
role of an effective temperature which contains information on the frozen
degrees of freedom. This temperature field introduces thermal barriers in the
system \cite{sastry}. The presence of these thermal barriers is a
nonequilibrium effect that disappears when the system is at equilibrium. Over
long time, when the transient regime has died out, Eq. (\ref{precurrent})
becomes%
\begin{align}
J(x,t)  &  =-\tau_{0}\{\rho(x,t)\left(  \nabla V(x)-f(t)\right)
\label{current}\\
&  +k_{B}\nabla\rho(x,t)T(x,t)\}\text{ }\nonumber
\end{align}
where $\tau_{0}=\gamma^{-1}$. \ 

On the other hand, in Eqs. (\ref{precurrent}) and (\ref{current}) the current
$J(x,t)$ appears coupled to the temperature field $T(x,t)$ which evolves
according to
\begin{gather}
\frac{\partial}{\partial t}\rho(x,t)T(x,t)=-\frac{2}{k_{B}}\nabla
\rho(x,t)h(x,t)\label{tempertaurebalance}\\
-2\left(  \nabla V(x)-f(t)\right)  \frac{J(x,t)}{k_{B}}-\frac{2}{k_{B}\tau
_{0}}\rho(x,t)\left(  T(x,t)-T_{o}\right)  \text{ ,}\nonumber
\end{gather}
obtained by multiplying Eq. (\ref{F-P}) by $u^{2}$ and integrating in $u$.
Here, we have defined the heat flow%
\begin{equation}
h(x,t)=\frac{1}{2}\int u^{3}\phi_{x}(u,t)du\text{ .} \label{heat}%
\end{equation}
For times $t\gg\tau_{0}$, Eq. (\ref{tempertaurebalance}) reduces to
\begin{gather}
\rho(x,t)k_{B}\left(  T(x,t)-T_{o}\right)  =-\tau_{0}\nabla\rho
(x,t)h(x,t)\label{temperature difference}\\
-\tau_{0}\left(  \nabla V(x)-f(t)\right)  J(x,t)\text{ .}\nonumber
\end{gather}
In the particular case of equilibrium, $\phi_{x}(u,t)$ would be an equilibrium
Maxwellian, thus $J(x,t)=h(x,t)=0$ which would lead to $T(x,t)=T_{o}$. If, on
the other hand, we integrate Eq. (\ref{temperature difference}) we obtain%
\begin{equation}
k_{B}\left(  T(t)-T_{o}\right)  =\tau_{0}\frac{dq}{dt}\text{ ,}
\label{tempertures}%
\end{equation}
with $T(t)=\left\langle T(x,t)\right\rangle $. This last result together with
Eq. (\ref{heatrate}) leads to the interesting conclussion that even if
$f(t)=0$ the partial elimination of the fast variable would introduce a
nonequilibrium temperature \cite{agusti2}. Only for the particular case of a
soft energy landscape with small gradients $\nabla V(x)$, the difference
$T(t)-T_{o}$ would be negligible in comparison to $T_{o}$. We can make this
statement more precise by defining a critical length scale $L=\triangle
V/\nabla V(x)$, where $\triangle V$ is the characteristic variation of the
potential, such that if $\left(  \gamma L\right)  ^{2}/k_{B}T_{o}\ll1$, then
$T(x,t)=$ $T_{o}$.

In fact the Brownian motion as described by the Klein-Kramers equation
(\ref{F-P}) is a problem with multiple time scales as was pointed out in refs.
\cite{barreiro}, \cite{bocquet} and implicitely included in the generalization
of the Smoluchowski equation found previously in \cite{wilenski},
\cite{titulaer}. Here, we have embedded all of these scales in the
nonequilibrium temperature $T(x,t)$ or $T(t)$, constituting a kind of mean
field approach which leads to the nonexponential relaxation and aging we will
show in the following sections.

To sum up, in this section we have obtained the evolution equations for the
first moments of the conditional probability density which are of interest for
the thermodynamic description of the Brownian gas. More detail and some
consequences will be shown in the following sections.

\section{Generalized diffusion equation}

Here, we will establish the generalized diffusion equation and examine its
consequences in respect to the relaxation phenomena.

After defining the effective potential $\Phi(x,t)=V(x)+k_{B}T(x,t)$, Eq.
(\ref{current}) can be expressed
\begin{align}
J(x,t)  &  =-D(x,t)\nabla\rho(x,t)-\tau_{0}\rho(x,t)\left(  \nabla
\Phi(x,t)-f(t)\right)  \text{ }\label{current2}\\
&  \equiv Q(x,t)+\tau_{0}\rho(x,t)f(t)\text{ ,}\nonumber
\end{align}
where $D(x,t)=\tau_{0}k_{B}T(x,t)$ is the generalized diffusion coefficient
and $Q(x,t)$ corresponds to the current of the purely relaxational system
(\textit{i.e. }in absence of nonconservative forces). \ Thus, by substituting
Eq. (\ref{current2}) into Eq. (\ref{continuity}) and assuming $\tau_{0}=1$
(\emph{i.e. }rescaling the time t), this last equation becomes the generalized
diffusion equation
\begin{equation}
\frac{\partial}{\partial t}\rho(x,t)+\nabla\text{ }\rho(x,t)f(t)=\nabla
Q(x,t)\text{ .} \label{diffusionequation}%
\end{equation}
In addition, as a consequence of the elimination of degrees of \ freedom, the
dynamics becomes non-Markovian and depends on the equilibrium temperature
$T_{0}$ at the quench time. In the frame of reference defined through the
transformation $x^{\ast}=x-\int^{t}f(v)dv$, the current (\ref{current2})
reduces to $Q(x^{\ast},t)$. Thus, in this frame one has a purely relaxational
process for which%
\begin{equation}
\frac{\partial}{\partial t}\rho(x^{\ast},t)=\nabla Q(x^{\ast},t)\text{ .}
\label{difusioneq2}%
\end{equation}
This equation admits a quasi-equilibrium solution $\rho_{qe}(x,t)\sim
\exp\left\{  -\int^{x}\frac{1}{k_{B}T(x^{^{\prime}},t)}\nabla^{^{\prime}}%
\Phi(x^{^{\prime}},t)dx^{^{\prime}}\right\}  $ for which $Q(x,t)=0$, where to
simplify the notation we have omitted the superscript $^{\ast}$ on the $x$.

\subsection{Population dynamics and relaxation function}

To examine the problem of the population dynamics with its origin in the
generalized diffusion equation (\ref{difusioneq2}) and to find the
corresponding relaxation function we will assume that at a point $x_{2}$ there
is a sink where $\rho(x_{2},t)=0$ and a source at $x_{1}$. Hence, a
quasi-stationary current, $Q(t)$ can be established in the system, thus,
integration of \ $Q(x,t)$\ gives us the quasi-stationary probability density
in terms of $Q(t)$
\begin{equation}
\rho(x,t)=Q(t)\rho_{qe}(x,t)\int_{x}^{x_{2}}\frac{dy}{D(y,t)\rho_{l.qe.}%
(y,t)}\text{ .} \label{density}%
\end{equation}
On the other hand, a second integration leads to
\begin{equation}
n(t)=Q(t)\int_{x_{1}}^{x_{b}}dx\rho_{qe}(x,t)\int_{x}^{x_{2}}\frac
{dy}{D(y,t)\rho_{l.qe.}(y,t)}\text{,} \label{density2}%
\end{equation}
with $n(t)=\int_{x_{1}}^{x_{b}}\rho(x,t)dx$ being the population at the left
of the sink. Thus, after integration of Eq. (\ref{continuity}) we obtain%
\begin{equation}
\frac{d}{dt}n(t)=-\left\{  Q(x_{b},t)-Q(x_{1},t)\right\}  =-Q(t)\text{ ,}
\label{partialrate}%
\end{equation}
which leads to the following rate equation
\begin{equation}
\frac{d}{dt}n(t)=-K(t;x_{b})n(t)\text{ ,} \label{rateequation}%
\end{equation}
where
\begin{equation}
K(t;x_{b})^{-1}=\int_{x_{1}}^{x_{b}}dx\rho_{qe}(x,t)\int_{x}^{x_{2}}\frac
{dy}{D(y,t;t_{0})\rho_{l.qe.}(y,t)} \label{rateconstant}%
\end{equation}
is the rate constant, and $x_{b}$ is the position of the barrier between
$x_{1}$ and $x_{2}$. The relaxation equation (\ref{rateequation}) admits the
solution
\begin{align}
n(t)  &  =n(t_{0})\exp\left\{  -\int_{t_{0}}^{t}K(t^{\prime};x_{b})dt^{\prime
}\right\} \label{population}\\
&  \equiv n(t_{0})\text{ }\exp\left\{  -g(t)\right\} \nonumber
\end{align}
or
\begin{equation}
\widehat{n}(t)=\widehat{n}(t_{0})\text{ }\exp\left\{  -g(t/\tau_{0})\right\}
\text{ ,} \label{population2}%
\end{equation}
where $\widehat{n}(t)=n(t/\tau_{0})$. Both Eqs. (\ref{population}) and
(\ref{population2}) constitute an interesting result since we obtain a
hierarchy of relaxation times starting from a Markovian equation, the
Fokker-Planck equation (\ref{F-P})\textbf{. }Here, the function $g(t/\tau
_{0})$ might be interpreted as an algebraic function $(t/\tau_{0})^{\beta}$
obtaining a stretched exponential behavior \cite{angell} or as a logarithmic
function $-\alpha\log At$ which leads to a power law behavior \cite{hu} that
characterizes anomalous diffusion \cite{metzler}. In addition, another reading
of our result is possible because assuming the form of $g(t/\tau_{0})$ is
equivalent to assuming the form of the distribution of residence times
$\psi(t)\equiv-\frac{d}{dt}n(t)$ \cite{gezelter}. Thus, an in-depth analysis
of the implications of the supposition of a nonexponential distribution of
residence times usually performed in Continuous Time Random Walk models,
reveals that this hypothesis might also be rooted in the nonequilibrium
character of the dynamics inherent to the energy landscape picture as we have shown.

\subsection{Hierarchy of relaxation times}

Inherent to the nonexponential relaxation Eqs. (\ref{population}) and
(\ref{population2}) there is a hierarchy of relaxation times. To show this
clearly here, we will derive the hierarchy of relaxation times for the
particular case of a bistable potential and the state of quasi-equilibrium.
Thus, let us assume here that $V(x)$ is a bistable potential with one well
much deeper than the other. In this case, we can obtain the dynamic of the
population of the shalower well through an equation analogous to Eq.
(\ref{rateequation}), where now according to the Kramers formula
\begin{equation}
K=f_{0}\exp\left\{  -\frac{\Delta}{k_{B}T(t)}\right\}  \text{ ,}
\label{frequency}%
\end{equation}
where we have defined the frequency $f_{0}=\frac{1}{2}\sqrt[2]{V^{\prime
\prime}(x_{1})\left\vert V^{\prime\prime}(x_{b})\right\vert }$, and the
barrier height $\Delta=V(x_{b})-V(x_{1})$. Thus,
\begin{equation}
n(t)=n(t_{0})\exp\left\{  -\int_{t_{0}}^{t}\frac{dt^{\prime}}{\tau(t^{\prime
})}\right\}  \text{ ,} \label{population3}%
\end{equation}
with
\begin{equation}
\tau(t)=\frac{1}{f_{0}}\exp\left\{  \frac{\Delta}{k_{B}T(t)}\right\}
\label{relaxationtime}%
\end{equation}
From Eq. (\ref{tempertures}), we can write
\begin{equation}
T(t)-T(t_{0})=f(t)^{2}-f(t_{0})^{2}\text{ ,} \label{temperaturediferences}%
\end{equation}
where now $T(t)=\left\langle T(x,t)\right\rangle _{qe}$ and $\left\langle
\nabla V(x)u\right\rangle _{qe}=0$. Therefore,
\begin{equation}
\tau(t)=\frac{1}{f_{0}}\exp\left\{  \frac{\Delta/k_{B}}{f(t)^{2}-f(t_{0}%
)^{2}+T(t_{0})}\right\}  \text{ ,} \label{relaxatio1}%
\end{equation}

or equivalently%
\begin{equation}
\tau(t)=\frac{1}{f_{0}}\exp\left\{  \frac{\Delta/k_{B}}{T^{\ast}(t)-T^{\ast
}(t_{0})}\right\}  \exp\left\{  \frac{1}{1+\frac{T(t_{0})}{T^{\ast}%
(t)-T^{\ast}(t_{0})}}\right\}  \text{ ,} \label{intermediated}%
\end{equation}
where $T^{\ast}(t)=T(t)-T_{o}$, \textit{i.e. }a scale of temperature whose
zero point coincides with the temperature of the heat bath. Here, assuming
that $\frac{T^{\ast}(t)-T^{\ast}(t_{0})}{T(t_{0})}\ll1$ we can neglect the
second exponential on the right hand side of Eq. (\ref{intermediated}),
leading to the result%
\begin{equation}
\tau(t)=\frac{1}{f_{0}}\exp\left\{  \frac{\Delta/k_{B}}{T^{\ast}(t)-T^{\ast
}(t_{0})}\right\}  \label{VTF}%
\end{equation}
similar to the celebrated Vogel-Tammann-Fulcher equation. In the next order of
approximation we would have%
\begin{equation}
\tau(t)=\frac{1}{f_{0}}\left(  \frac{T^{\ast}(t)-T^{\ast}(t_{0})}{T(t_{0}%
)}\right)  \exp\left\{  \frac{\Delta/k_{B}}{T^{\ast}(t)-T^{\ast}(t_{0}%
)}\right\}  \text{ } \label{firstorder}%
\end{equation}
providing an expression for the relaxation time which is finite when $T^{\ast
}(t)=T^{\ast}(t_{0})$.

\section{Fluctuation-dissipation theorem}

Here, we want to derive the fluctuation-dissipation theorem for the
nonequilibrium system, which is locally satisfied and provides us with a way
to compute the nonequilibrium temperature. In order to highlight the main
details and avoid lengthy calculations, we will homogenize Eq.
(\ref{difusioneq2}) introducing a preaveraging condition with respect to the
quasi-equilibrium distribution
\begin{equation}
\frac{\partial}{\partial t}\rho=\frac{\partial}{\partial x}\left\{
D(t)\frac{\partial}{\partial x}\rho(x,t)+\rho(x,t)\frac{\partial}{\partial
x}\Phi(x,t)\right\}  \text{ ,} \label{smoluchowski}%
\end{equation}
where $D(t)=k_{B}T(t)$ is the preaveraged diffusion coefficient. Thus, now the
quasi-equilibrium density becomes a local Boltzmann probability density
$\rho_{qe}(x,t)\sim\exp\left\{  -\Phi(x,t)/k_{B}T(t)\right\}  $ .

Consider a time-dependent external field $F(t)=F\Theta(t-t_{w})$ applied to
the system Eq. (\ref{smoluchowski}) at the waiting time $t_{w}$ when this
system is in the quasi-equilibrium state described by $\rho_{qe}$. Here,
$\Theta(t-t_{w})$ is the step function. This force modifies the potential, and
we assume that the perturbed potential is $\widetilde{\Phi}(x,t)=\Phi
(x,t)-F(t)B(x)$. If the field is weak, the deviation with respect to
stationarity of the average value of any function $A(x)$ of the configuration
of the system is a linear functional of the field
\begin{equation}
\left\langle A(t)\right\rangle _{F}-\left\langle A(t)\right\rangle _{qe}%
=\int_{t_{w}}^{t}R(t,t^{\prime})F(t^{\prime})dt^{\prime}\text{ ,}
\label{response}%
\end{equation}
which introduces the response function $R(t,t^{\prime})$. Now $\left\langle
A(t)\right\rangle _{F}$ can be calculated once the perturbed distribution
function $\widetilde{\rho}(x,t)$ is known:
\begin{equation}
\left\langle A(t)\right\rangle _{F}=\int A(x)\widetilde{\rho}(x,t)dx\text{ .}
\label{average}%
\end{equation}
Since the system is in the nonequilibrium stationary state at the time $t_{w}%
$, $\widetilde{\rho}(x,t)$ is related to $\rho_{qe}(x,t_{w})$ thrugh the
perturbed Green function $\widetilde{G}(x,t\mid x^{\prime},t_{w})$%
\begin{equation}
\widetilde{\rho}(x,t)=\int\widetilde{G}(x,t\mid x^{\prime},t_{w})\rho
_{qe}(x^{\prime},t_{w})dx^{\prime}\text{ ,} \label{green}%
\end{equation}
and
\begin{equation}
\widetilde{\rho}_{qe}(x,t)=\frac{\exp\left(  FB(x)/k_{B}T(t)\right)
}{\left\langle \exp\left(  FB(x)/k_{B}T(t)\right)  \right\rangle _{qe}}%
\rho_{qe}(x,t)\text{ .} \label{perturbedequilibrium}%
\end{equation}
To compute $\left\langle A(t)\right\rangle _{F}$ by using Eq. (\ref{average})
with the help of Eq. (\ref{green}) we have to know the relation between
$\widetilde{G}(t,x\mid x^{\prime},t_{w})$ and the corresponding unperturbed
Green function $G(t,x\mid x^{\prime},t_{w})$. This relation can be established
from the identity
\begin{equation}
\widetilde{\rho}_{qe}(x,t)=\int\widetilde{G}(t,x\mid x^{\prime},t_{w}%
)\widetilde{\rho}_{qe}(x^{\prime},t_{w})dx^{\prime}\text{ , }
\label{equilibriumidentity}%
\end{equation}
by applying the relation (\ref{perturbedequilibrium}) at both sides of Eq.
(\ref{equilibriumidentity}), leading to
\begin{equation}
G(t,x\mid x^{\prime},t_{w})=\frac{\exp\left(  FB(x)/k_{B}T(t_{w})\right)
}{\exp\left(  FB(x)/k_{B}T(t)\right)  }\widetilde{G}(t,x\mid x^{\prime}%
,t_{w})\frac{\left\langle \exp\left(  FB(x)/k_{B}T(t)\right)  \right\rangle
_{qe}}{\left\langle \exp\left(  FB(x)/k_{B}T(t_{w})\right)  \right\rangle
_{qe}}\text{ .} \label{propagators}%
\end{equation}
By using Eqs. (\ref{average}), (\ref{green}) and (\ref{propagators}), after
expanding the exponentials with respect to F, one then obtains
\begin{equation}
\left\langle A(t)\right\rangle _{F}-\left\langle A(t)\right\rangle
_{qe}=F\left\{  \frac{1}{k_{B}T(t)}C_{A,B}(t,t)-\frac{1}{k_{B}T(t_{w})}%
C_{A,B}(t,t_{w})\right\}  \text{ ,} \label{response2}%
\end{equation}
where $C_{A,B}(t,t^{\prime})=\left\langle A(t)B(t^{\prime})\right\rangle
-\left\langle A(t)\right\rangle \left\langle B(t^{\prime})\right\rangle $. We
assume that the state of the system varies slowly enough so that it is safe to
replace $T(t)$ for $T(t_{w})$. Thus, one achieves%
\begin{equation}
\left\langle A(t)\right\rangle _{F}-\left\langle A(t)\right\rangle _{qe}%
=\frac{1}{k_{B}T(t_{w})}\int_{-\infty}^{t}\frac{\partial}{\partial t^{\prime}%
}C_{A,B}(t,t^{\prime})F(t^{\prime})dt^{\prime}\text{ ,}
\label{responsefunction}%
\end{equation}
which defines the nonequilibrium response function
\begin{equation}
R(t,t_{w})=\frac{1}{k_{B}T(t_{w})}\frac{\partial}{\partial t_{w}}%
C_{A,B}(t,t_{w})\text{ ,} \label{nonequilibriumresponse}%
\end{equation}
for $t>t_{w}$, constituting a local version of the fluctuation-dissipation
theorem, the GFDT.

This section shows that the kind of effects related to the nonequilibrium
nature of the solutions araised in the Fokker-Planck equation can be tested by
means of linear response theory.

\section{Conclusions}

Here, we have learned that the dynamics of a brownian gas under the action of
a nonconservative force slowly varying over time is representative of the
complex dynamics observed in dissipative systems. After quenching, the system
is taken out of its initial equilibrium and then a heat flow is established.
We model the effects of the quench of the system through the elimination of
the fast degrees of freedom from the dynamics described by the Fokker-Planck
equation. The departure from equilibrium and consequently the exchange of heat
is maintained by means of the nonconservative force. We have also considered
the existence of a nonperiodic ruogh potential in order to incorporate
activated processes into our analysis according to the energy landscape
paradigm inherent to several complex systems. A series of facts constituting
the main results of our analysis follows.

First, the frozen degrees of freedom are manifested in the constrained
dynamics through an effective temperature field different from the equilibrium
temperature. The average of this temperature field plays the role of the
temperature that characterizes the non-Markovian diffusion equation obtained
from the Fokker-Planck equation, which describes the dynamic of the slow
degrees of freedom . We also derive the relaxation equation for the effective
temperature field from the Fokker-Planck equation which enables us to give a
sound base to the nonequilibrium temperature as a result of the balance
between the added heat and the heat generated during the relaxation process in
the system. In addition, we show that the nonequilibrium temperature computed
using the quasi-equilibrium distribution, \textit{i.e.} the stationary
solution of the non-Markovian diffusion equation, characterizes the
generalized fluctuation-dissipation theorem we have derived, thus coinciding
with the effective temperature referred to by several authors in the
literature \cite{agusti}, \cite{kurchan}.

On the other hand, we find a hierarchy of relaxation times as a consequence of
the non-Markovian character of the constrained dynamics of the system. In this
respect, our model also describes stretched exponential relaxation as well as
anomalous diffusion \cite{hu}, \cite{metzler} and other Continuous Time Random
Walk problems \cite{gezelter}. Likewise, as a consequence of the
nonexponential relaxation, we obtain an equation similar to the celebrated
Vogel-Tammann-Fulcher law which is, in our opinion, an important result.

Other approaches based on the energy landscape, considering the relaxation as
a series of activated processes, assume that the activation energies are
independent from the temperature but with a certain random spatial
distribution. Averaging over this distribution of energy barrier gives one the
corresponding relaxation \cite{ediger}, \cite{berne}, \cite{wolynes2}. In the
literature another types of models undergoing nonexponential relaxation can be
found, as for example, disorder models \cite{klafter}.

We conclude that our model incorporating the multiplicity of time scales
embedded in the Fokker-Planck equation, although simple, contains the main
features in the dynamical slowing down observed in a wide variety of complex
systems. On the other hand, with our approach we reach more insight avout the
physical meaning of the corrections to the Smoluchowski equation found in
previous works as those referred to in this contribution, stating this problem
in the modern language of the complex dynamics. In addition, our model
constitutes a way to study Brownian motion in disordered space leading to
results similar to those reached through the approach based on fractal
Fokker-Planck equations . Therefore, we think this work will contribute to the
understanding of some aspects of complex dynamics in several dissipative systems.

\begin{acknowledgement}
This work was supported by DGICYT of the Spanish Government under Grant No. PB2002-01267.
\end{acknowledgement}


\begin{thebibliography}{99}                                                                                               %


\bibitem {ngai}C. A. Angell, K. L. Ngai, G. B. McKenna, P. F. McMillan and S.
W. Martin, J. Appl. Phys. \textbf{88}, 3113 (2000).

\bibitem {wolynes1}J.D. Bryngelson and P.G. Wolynes, Proc. Natl. Acad. Sci.
USA \textbf{84}, 7524 (1987).

\bibitem {sabelko}J. Sabelko, J. Ervin, and M. Gruebele, Proc. Natl. Acad.
Sci. USA \textbf{96}, 6031 (1999).

\bibitem {shlesinger}M.F. Shlesinger, G.M. Zaslavsky and J. Klafter, Nature
\textbf{363}, 31 (1993).

\bibitem {balescu}R.Balescu, Cond. Matt. Phys. \textbf{1}, 815 (1998).

\bibitem {stillinger}P.G. Debenedetti and F. H. Stillinger, Nature
\textbf{410}, 259 (2001).

\bibitem {wolynes}H. Frauenfelder, S. Sligar and P. G. Wolynes, Science
\textbf{254}, 1598 (1991).

\bibitem {sastry}S. Sastry, P.G. Debenedetti, and F.H. Stillinger, Nature
\textbf{393}, 554 (1998).

\bibitem {ediger}M.D. Ediger, Annu. Rev. Phys. Chem. \textbf{51}, 99 (2000).

\bibitem {inertial}J.M. Rub\'{\i}, A. P\'{e}rez-Madrid, Physica A
\textbf{264}, 492 (1999).

\bibitem {mesoscopic}P. Mazur, Physica A \textbf{274}, 491 (1999).

\bibitem {hill}T.L. Hill, \textit{Thermodynamics of Small Systems} (Benjamin,
New York, 1963). Part I.

\bibitem {vankampen}N.G. van Kampen, \textit{Stochastic Processes in Physics
and Chemistry }(North-Holland, Amsterdam, 1990), Sect. VII.6, pag. 202.

\bibitem {degroot}S.R. de Groot and P. Mazur, \textit{Non-Equilibrium
Thermodynamics }(Dover, New York, 1984), Sect. VII.8, pag 126.

\bibitem {ivan}I. Santamaria-Holek, A. Perez-Madrid, J. M. Rubi, J. Chem.
Phys. \textbf{120}, 2818 (2004).

\bibitem {agusti2}A. P\'{e}rez-Madrid, Phys. Rev. E \textbf{69, }062102 (2004).

\bibitem {barreiro}L.A. Barreiro, J.R. Campanha and R.E. Lagos, Physica A
\textbf{283}, 160 (2000).

\bibitem {bocquet}L. Bocquet, Am. J. Phys. \textbf{65, }140 (1997).

\bibitem {titulaer}U.M. Titulaer, Physica A \textbf{91}, 321 (1978).

\bibitem {wilenski}G. Wilenski, J. Stat. Phys. \textbf{14, }153 (1976).

\bibitem {agusti}A. P\'{e}rez-Madrid, D. Reguera, J.M. Rub\'{\i}, Physica A
\textbf{329}, 357 (2003).

\bibitem {kurchan}L.F. Cugliandolo, J. Kurchan, L. Peliti, Phys. Rev. E
\textbf{55}, 3898 (1997).

\bibitem {angell}C.A. Angell, K.L. Ngai, G.B. McKenna, P.F. McMillan, and S.W.
Martin, J. Appl. Phys. \textbf{88}, 3113 (2000).

\bibitem {hu}Hu Cang, Jie Li, V.N. Novikov, and M.D. Fayer, J. Chem. Phys.
\textbf{118}, 9303 (2003).

\bibitem {metzler}R. Metzler and J. Klafter, Phys. Rep. \textbf{339}, 1 (2000).

\bibitem {gezelter}C.F. Vardeman II and J.D. Gezelter, J. Phys. Chem. A
\textbf{105}, 2568 (2001).

\bibitem {berne}E. Rabani, J.D. Gezelter, and B.J. Berne, Phys. Rev. Lett.
\textbf{82}, 3649 (1999).

\bibitem {wolynes2}X. Xia and P.G. Wolynes, Phys. Rev. Lett. \textbf{86}, 5526 (2001).

\bibitem {klafter}J. Klafter and M.F. Shlesinger, Proc. Natl. Acad. Sci.
U.S.A. \textbf{83}, 848 (1986).
\end{thebibliography}
\end{document}